\documentclass[runningheads]{llncs}
\usepackage[T1]{fontenc}
\usepackage{graphicx}
\usepackage{orcidlink}
\usepackage{tikz}
\usepackage{listings}
\usepackage{todonotes}
\usepackage{subcaption}
\usepackage{lstautogobble}
\usepackage{xspace}
\usepackage{cleveref}
\usepackage{ifthen}
\usepackage[inline]{enumitem}
\usepackage{booktabs}
\usepackage{marvosym}

\usetikzlibrary{backgrounds,shapes.symbols,positioning,fit,calc,arrows.meta,shapes.geometric}

\tikzset{
  ->/.style={-{Stealth}, shorten >=0.1mm},
  <-/.style={{Stealth}-, shorten <=0.1mm},
  rect/.style={draw,fill=teal!20,thick},
  rrect/.style={rect,rounded corners=2mm}
}

\definecolor{bluekeywords}{rgb}{0.13, 0.13, 1}
\definecolor{greentypes}{rgb}{0, 0.5, 0}
\definecolor{orangecomments}{rgb}{1, 0.5, 0.1}
\definecolor{redstrings}{RGB}{171, 114, 2}
\definecolor{graynumbers}{rgb}{0.5, 0.5, 0.5}
\definecolor{goldcomments}{rgb}{0.6, 0.4, 0.08}

\definecolor{implemented}{rgb}{0.67, 0.9, 0.93}
\colorlet{existing}{lightgray}

\lstdefinelanguage{Lola}{
  keywords=[0]{input, output, trigger, constant, import, spawn, eval, close, with, when},
  moredelim=**[is][\transparent{0.6}]{?}{?},
  moredelim=**[is][\color{greentypes}@]{@}{@},
  keywordstyle=[0]\bfseries\color{bluekeywords},
  keywords=[1]{if, then, else, aggregate, defaults, offset, last, by, or, to, sin, cos, abs, hold, over, using, over_instances, prob, query, query_unchecked, as, into},
  keywords=[2]{Variable, String, Int, Int64, UInt, UInt64, Bool, Float32, Float64, Float, QueryId, QueryStatus},
  keywordstyle=[2]\color{greentypes},
  sensitive=false,
  comment=[l]{//},
  morecomment=[s]{/*}{*/},
  morestring=[b]',
  morestring=[b]",
  literate={\\@}{@}1,
  moredelim=[s][\color{purple}\bfseries]{\#[}{]},
}
\makeatletter
\newcommand{\rtlola}{\textsc{RTLola}\xspace}

\newboolean{fullversion}
\setboolean{fullversion}{true}

\begin{document}

\title{Extending RTLola with External Data Queries}

\author{
Bernd Finkbeiner\inst{1,2}\orcidlink{0000-0002-4280-8441} \and
Jakob Hirschler\inst{3} \and\\
Frederik Scheerer\textsuperscript{(\Letter)}\inst{1}\orcidlink{0009-0007-8115-0359} \and
Sebastian Schirmer\inst{3}\orcidlink{0000-0002-4596-2479}
}
\authorrunning{Finkbeiner et al.}

\institute{
    CISPA Helmholtz Center for Information Security, Saarbrücken, Germany \\
    \email{\{finkbeiner, frederik.scheerer\}@cispa.de} \and
    Technical University of Munich, Germany 
    \and
    German Aerospace Center (DLR), Braunschweig, Germany \\
    \email{\{jakob.hirschler, sebastian.schirmer\}@dlr.de}
}

\maketitle

\begin{abstract}
Stream-based monitoring enables the concise specification of complex temporal properties.
However, existing stream-based monitors are limited when dealing with large external data sources, a task that is better handled by specialized data management systems.
We address these limitations by extending stream-based monitors with the ability to query external data sources.
We implement this approach in \rtlola and investigate challenges such as handling delayed responses, type checking of returned data, and runtime error management.
A unified interface enables the seamless integration of existing systems into our approach, such as static databases or dynamic endpoints, e.g. a weather API.
Our evaluation using specifications from the aviation domain also shows that a custom geospatial backend based on k-d trees outperforms state-of-the-art database systems.

\end{abstract}

\section{Introduction}\label{sec:intro}
Runtime monitoring is particularly important for safety-critical applications such as aviation, where violations of safety requirements may have severe consequences and timely reactions are needed.
In practical monitoring scenarios, specifications often depend not only on the sensor data produced by the monitored system itself, but also on large external data sources.
These sources include static geospatial data, such as terrain or obstacle information, as well as dynamically changing information, such as live weather data.
One possible approach is to store all static information directly within the monitor, for example, obstacle locations used to determine which obstacles are currently close to the aircraft.
However, stream-based monitoring systems are designed for the efficient evaluation of temporal properties over streams and are not optimized for storing and querying large amounts of data.
In other situations, such as retrieving live weather information, the required data is not even available when the monitor is initialized and must instead be obtained dynamically through external services.
We therefore propose storing such information in dedicated external systems specifically designed and optimized for these tasks, and enabling the monitor to actively query these systems whenever additional information is required during runtime.

Querying external information from the monitor raises several challenges:
\begin{enumerate}
    \item Queries are asynchronous and may exhibit significant latency.
    During this time, the monitor must not block and must continue the evaluation of unrelated monitoring tasks.
    \item Queries may fail due to unavailable services or connectivity issues.
    The monitor must not crash and must allow the specification to detect when queries fail and define suitable fallback behavior.
    \item The data format expected and returned by the queried system must be validated to ensure type safety and prevent runtime errors.
    \item Different monitoring applications require querying a wide variety of backends, including general-purpose relational databases, specialized geospatial database systems, and external APIs.
    All of which may use different querying languages, communication protocols, and data formats.
\end{enumerate}

\begin{figure}[t]
    \centering
    \begin{tikzpicture}
    \node[rrect,minimum width=2.5cm,minimum height=2cm,text depth=0.8cm,] (monitor) {\large Monitor};

    \node[rrect,minimum width=1.8cm,minimum height=1.5cm,align=center,right=0.5cm of monitor] (interface) {Query\\Interface};

    \foreach \i in {0,...,3}{
        \draw[<-,green!60!black] ([xshift=\i*3mm,xshift=4mm]monitor.north west) -- ++(0,7mm);
    }
    \foreach \i in {0,...,5}{
        \draw[->,blue] ([xshift=\i*4mm,xshift=2.5mm]monitor.south west) -- ++(0,-5mm);
    }
    \foreach \i/\s in {0/B,1/A}{
        \draw[->,transform canvas={yshift=\i*3mm,yshift=-6mm}] (monitor) node[right=-1mm] {$\mathit{query}_\s$} -- (interface) ;
    }

    \draw[->,green!60!black] ([xshift=2mm]interface.north) node[above left=0mm and 4mm] {$\mathit{res}_A$} -- ++(0,9mm) -| ([xshift=-1mm,xshift=7mm]monitor);
    \draw[->,green!60!black] ([xshift=-2mm]interface.north) node[above right=0mm and 4mm] {$\mathit{res}_B$} -- ++(0,6mm) -| ([xshift=2mm,xshift=7mm]monitor);
    
    \node[left=2mm of monitor.north west,align=right,yshift=4mm,green!60!black] {Input\\Streams};
    \node[left=2mm of monitor.south west,align=right,yshift=-1.5mm,blue] {Output\\Streams};
    
    \node[right=7mm of interface,cylinder,rect,shape border rotate=90,aspect=0.2,inner sep=2mm,fill=lightgray!50] (database) {Database};
    \draw[->,transform canvas={yshift=3mm}] (interface) -- (database);
    \draw[<-,transform canvas={yshift=-1mm}] (interface) -- (database);
\end{tikzpicture}
    \caption{Overview of query extension for stream-based monitors.}
    \label{fig:overview}
\end{figure}
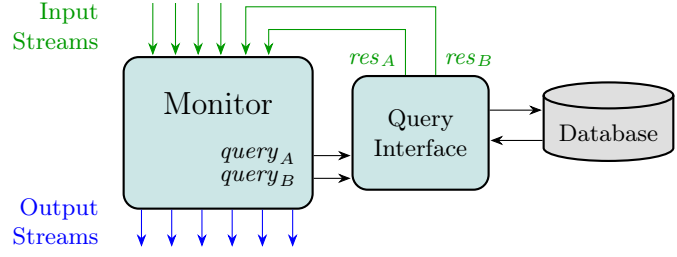

In this paper, we present our approach for integrating external data sources into stream-based runtime monitoring.
The overall architecture is illustrated in \Cref{fig:overview}.
As shown in the figure, queries are issued directly from the monitor as part of stream equations.
These queries are not executed synchronously, but are forwarded to a query interface, which acts as a communication layer between the monitor and external systems.
The interface forwards queries to an external backend, for example, a database, and asynchronously returns responses back to the monitor through dedicated input streams.
We have demonstrated our approach in an experimental evaluation using \rtlola~\cite{DBLP:conf/fm/BaumeisterFKS24}.
\rtlola's pacing type system allows us to precisely define when new queries are triggered.
To associate responses with the originating requests, each query returns a unique identifier, which is part of the results arriving at the inputs.
The implemented interface is general and extensible, abstracting away the complexities of asynchronous external queries.
In particular, it automatically handles issuing the queries according to the specification, associating the responses with the query identifier as well as with the associated inputs, and error management, while encapsulating backend-specific communication and serialization details.
We show the effectiveness of our approach on multiple monitoring scenarios from the aviation domain.

The rest of this paper is structured as follows:
First, we introduce the \rtlola query extension in \Cref{sec:contribution}. 
Then, we present implementation details and experimental results in  \Cref{sec:implementation}.

\subsection{Related Work}
Monitoring systems often require spatial and temporal reasoning.
This motivated logics like STREL~\cite{DBLP:conf/memocode/BartocciBLN17,DBLP:journals/lmcs/NenziBBL22}, SaSTL~\cite{DBLP:conf/iccps/MaBLS020}, SSTL~\cite{DBLP:conf/valuetools/NenziB14,DBLP:conf/rv/NenziBCLM15}, and more recently STL-GO~\cite{DBLP:journals/tecs/ZhaoYHFDL25}.
These extend temporal logic by operators over spatial objects, but do not address the underlying data sources, limiting scalability. 
We extend \rtlola to use external data sources that efficiently store and retrieve only relevant data.

\rtlola~\cite{DBLP:conf/fm/BaumeisterFKS24} is a stream-based language similar to TeSSLa \cite{kallwies2022tessla} and Striver \cite{gorostiaga2018striver}.
Unlike logics, it allows computations over input and output streams, enabling spatial computation such as the distance between two objects without introducing new operators.
Traditionally, such languages rely on external components to provide all relevant input streams.
In contrast, our approach follows an active monitoring paradigm \cite{DBLP:conf/rv/BaumeisterFS25}, where the monitor actively queries external data sources to obtain the information needed.

Although all objects could be encoded as constant streams in stream languages, specialized systems like geospatial databases \cite{postgis,tile38,DBLP:journals/ijgi/BreunigBJKMRASJ20} are more efficient for storing and processing large-scale data.
Instead of reimplementing these within the monitor, we directly leverage them.
At the same time, we provide essential guarantees for safety-critical applications, including delayed responses, type-safe processing of returned data, and runtime error management.

\section{Query-Extension to RTLola}\label{sec:contribution}

\rtlola~\cite{DBLP:conf/fm/BaumeisterFKS24} is a stream-based monitoring language that has been applied in domains including network monitoring \cite{faymonville2016stream}, aviation systems \cite{DBLP:conf/cav/BaumeisterFSST20,DBLP:conf/cav/BaumeisterFKLMST24}, automotive applications \cite{biewer2021rtlola} and algorithmic fairness~\cite{DBLP:conf/tacas/BaumeisterFSSW25,fairmon}.
Stream-based specifications define streams, which represent an infinite sequence of values.
\emph{Input streams} capture data produced by the monitored system, and \emph{output streams} are defined through stream equations, filtering and aggregating the data.
\emph{Triggers} are special, boolean-valued, output streams that specify violations of the specification: whenever a trigger evaluates to true, the specification is considered violated.
Consider the following specification from the aviation domain as an example.

\begin{lstlisting}
input pos : (Float64, Float64)
constant obstacle_lat: Float64 := 249.301
constant obstacle_lon: Float64 := 23.453
output distance := $\sqrt{(\mathtt{obstacle\_lat} - \mathtt{pos.0})^2 + (\mathtt{obstacle\_lon} - \mathtt{pos.1})^2}$
output closer := distance < distance.last(or: distance)
trigger closer $\land$ distance < 0.1 "Too close to the obstacle"
\end{lstlisting}

The specification defines one input stream, \lstinline!pos!, representing the current position of the aircraft.
Whenever a new GPS sensor reading arrives, a new tuple is added to this input stream.
Based on these inputs, the output stream \lstinline!distance! computes the distance between the current aircraft position and a fixed obstacle.
Another output stream, \lstinline!closer!, evaluates whether the newly computed distance is smaller than the previous one.
Finally, a trigger issues a warning whenever the aircraft moves closer to the obstacle while the current distance is below 0.1.
For a more detailed introduction to \rtlola based on exactly this scenario, we refer the reader to the \rtlola Tutorial~\cite{DBLP:conf/fm/BaumeisterFKS24}.

While the specification above successfully monitors a single obstacle, real-world scenarios typically require keeping track of a large number of obstacles simultaneously.
To handle this efficiently, in the following, we present our query extension, which addresses this challenge using the drone scenario as a running example.
Since obstacle information is static and potentially very large, it is maintained in an external geospatial database.
The monitor periodically queries this database to retrieve obstacles close to the current drone position.
The specification in this section is intentionally simplified; more realistic specifications illustrating the interaction of temporal operators with database queries are presented in the evaluation section.

Consider the following specification, extending the example above to monitor multiple obstacles, and counting the number of nearby ones.
\begin{lstlisting}
input pos : (Float64,Float64) as (lat,lon)
input obstacle : (QueryId,(Float64,Float64)) as (res_id,(o_lat,o_lon))
output obstacle_queries : QueryId @10s@ :=
    query("NEARBY obstacles POINT ?1 ?2 100",
           with: pos.hold(or: (0.0, 0.0)), into: obstacle)
output o_count @10s@ := obstacle.aggregate(over: 10s, using: count)
trigger o_count > 3 "Too many obstacles"
\end{lstlisting}
We initially assume that every query returns all results within ten seconds, so that queries can be processed independently without overlapping or interleaving results.
In addition to the \lstinline!pos! stream (Line 1), the specification defines an additional input stream \lstinline!obstacle! (Line 2), which receives the results of issued queries.
Notice the \lstinline!as! expression in the stream declaration, which introduces syntactic sugar that allows tuple values to be destructured and named directly.
For example, the latitude component of an obstacle can be accessed via \lstinline!o_lat!.
The output stream \lstinline!obstacle_queries! (Line 3) is evaluated periodically every \lstinline!@10s@!.
Whenever the \lstinline!query! expression is evaluated, a new query is issued to the database backend.
In this example, the backend is the geospatial database Tile38~\cite{tile38}, and the query string therefore corresponds to a Tile38 command.
While the query string itself is static, it contains placeholders \texttt{?1} and \texttt{?2}, which are instantiated at runtime through the \lstinline!with! expression (Line 5).
In this example, the query retrieves all obstacles within 100m around the current drone position (Line 4).
As specified by the \lstinline!into! clause (Line 5), all results are streamed to the \lstinline!obstacle! input stream. 
Finally, an output stream counts the obstacles returned in the last ten seconds (Line 6), which is then checked by a trigger (Line 7).

To address type safety, our extension provides two variants of the query expression.
The \lstinline!query! expression statically checks that the query result matches the declared type of the \lstinline!into! stream and that the placeholders are consistent with the \lstinline!with! expression, rejecting the specification at compile time if either check fails.
Where such static checking is not possible, for example if the backend cannot report its schema in advance, the unchecked variant \lstinline!query_unchecked! can be used instead.
In this case, the user must handle runtime errors in the specification directly.

Note, however, that the current specification relies on the assumption that all query results arrive within 10 seconds and that a new query is only issued after the previous interval has completed.
In time-critical domains, such a coarse interval is often insufficient.
Furthermore, in the specification above, the monitor can only reason about the number of nearby obstacles once the entire 10-second interval has elapsed, even if the query itself completes within a fraction of a second.
To address this, we introduce \emph{query identifiers}, which allow issued queries to be associated with the arriving results.
Each evaluation of the \lstinline!query! expression returns a unique identifier, and every returned result is annotated with the identifier of the query that produced it.
This allows the monitor to correlate results with their originating query regardless of when they arrive, decoupling the issuance of queries from the arrival of their results.
In particular, responses may arrive out of order, e.g., a query issued later may return before an earlier one, since each response is matched to its originating query independently via its identifier rather than by arrival order.
Using this mechanism, results can be processed independently of timing assumptions, allowing us to define a specification that counts the results associated with each issued query:
\begin{lstlisting}
output count(qid)
    spawn with obstacle_queries
    eval @obstacle@ when result_id = qid with count(qid).last(or: 0)+1
\end{lstlisting}
This specification introduces a \emph{parameterized stream}, consisting of multiple stream instances identified by the parameter \lstinline!qid!.
The \lstinline!spawn! expression determines when new instances are created.
In this example, a new instance is spawned for every query identifier produced by the \lstinline!obstacle_queries! stream.
The \lstinline!eval! expression is evaluated whenever a new value arrives on the \lstinline!obstacle! stream.
A new output value is produced only if the identifier attached to the result matches the identifier of the corresponding stream instance.
Each stream instance, therefore, maintains an independent count of the results of its associated query.

While query identifiers allow results to be associated with their corresponding queries, the monitor still lacks information about when a query is completed.
To resolve this, we introduce a dedicated status stream that communicates status information for each query in the specification.
Because this stream is shared between all queries, it is not associated with a query function, but instead globally annotated using \rtlola's annotation mechanism~\cite{DBLP:conf/rv/BaumeisterFS25}.
The backend reports successful completion or query failures over this input stream:
\begin{lstlisting}
#[query_status]
input status : (QueryId,QueryStatus) as (status_id,status)
trigger(qid)
    spawn with obstacle_queries
    eval @status@ status_id = qid $\land$ status = QueryStatus::Finished
        $\land$ count_per(qid).hold(or: 0) > 3 with "Too many obstacles"
    close @status@ when status_id = qid
\end{lstlisting}
The specification defines a parameterized trigger that is instantiated for every issued query.
Whenever a status update arrives, the trigger checks whether the corresponding query has completed successfully.
Only after completion does the specification check the number of obstacles from that query.
Further, all stream instances concerning a given identifier can be closed whenever a query finishes.

\section{Implementation and Experiments}\label{sec:implementation}

This section presents our implementation and evaluation with specifications from the aviation domain.
\ifthenelse{\boolean{fullversion}}{
    All specifications can be found in \Cref{app:specs}.
}{
    All specifications can be found in the full version~\cite{fullversion} of this paper.
}

\subsection{Implementation}

Our implementation extends the \rtlola StreamIR interpreter~\cite{DBLP:conf/cav/BaumeisterCFS25} with a query interface and four backend integrations.
The entire framework is implemented in Rust and uses \texttt{crossbeam} channels for non-blocking communication between the monitor and backend threads, enabling concurrent query execution without blocking the monitor.

\paragraph{Interface.}
To support a wide range of external systems, we provide a unified query interface that abstracts from backend-specific details, enabling the integration of databases and external APIs with minimal effort.
The interface exposes four methods.
$\mathit{openDatabase}$ establishes a connection to the backend.
$\mathit{prepareQuery}$ is called once per query statement before monitoring begins, allowing backends to create prepared statements and validate result types against the declared streams.
$\mathit{runQuery}$ is executed whenever a query is evaluated during monitoring.
It receives the current parameter values and two callback channels for streaming results and reporting status updates asynchronously.
Finally, $\mathit{finish}$ is called on monitor termination.

\subsection{Experiments}

\begin{figure}[t]
    \begin{subfigure}{0.49\linewidth}
    \includegraphics[width=\linewidth]{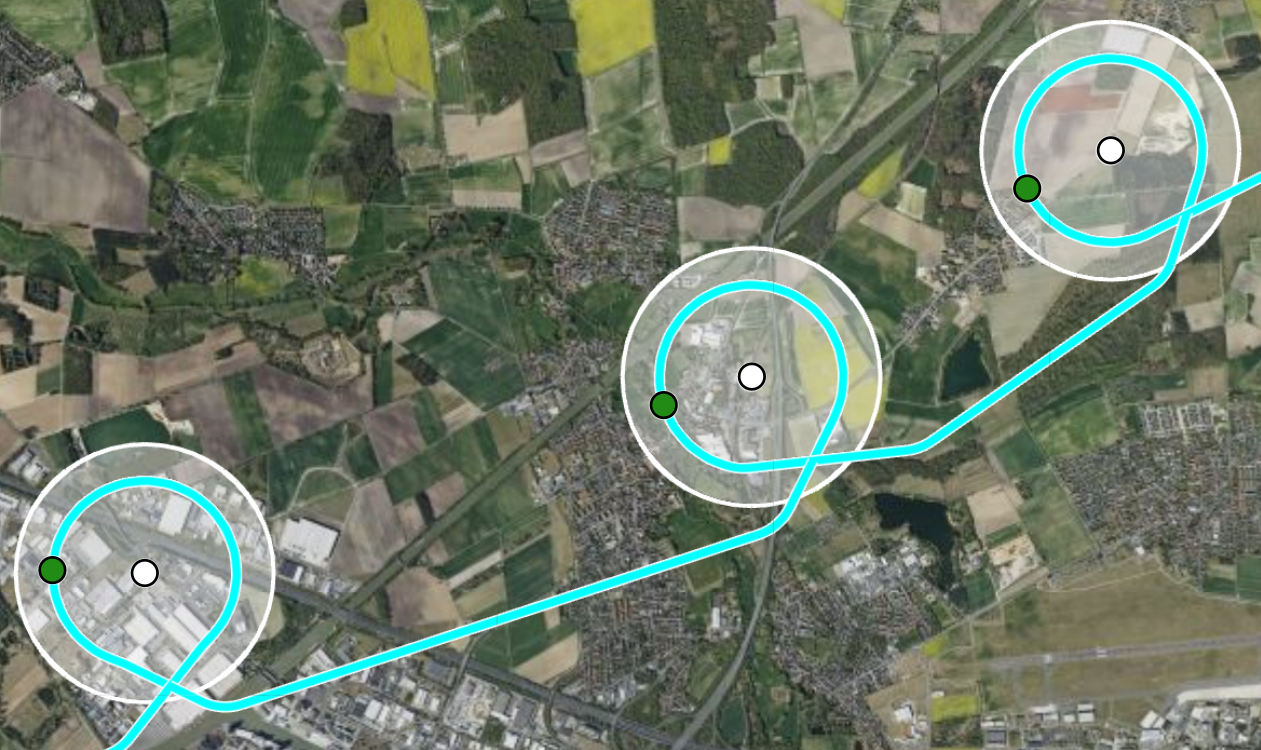} 
    \caption{A valid flight}
    \end{subfigure}
    \hfill
    \begin{subfigure}{0.49\linewidth}
    \includegraphics[width=\linewidth]{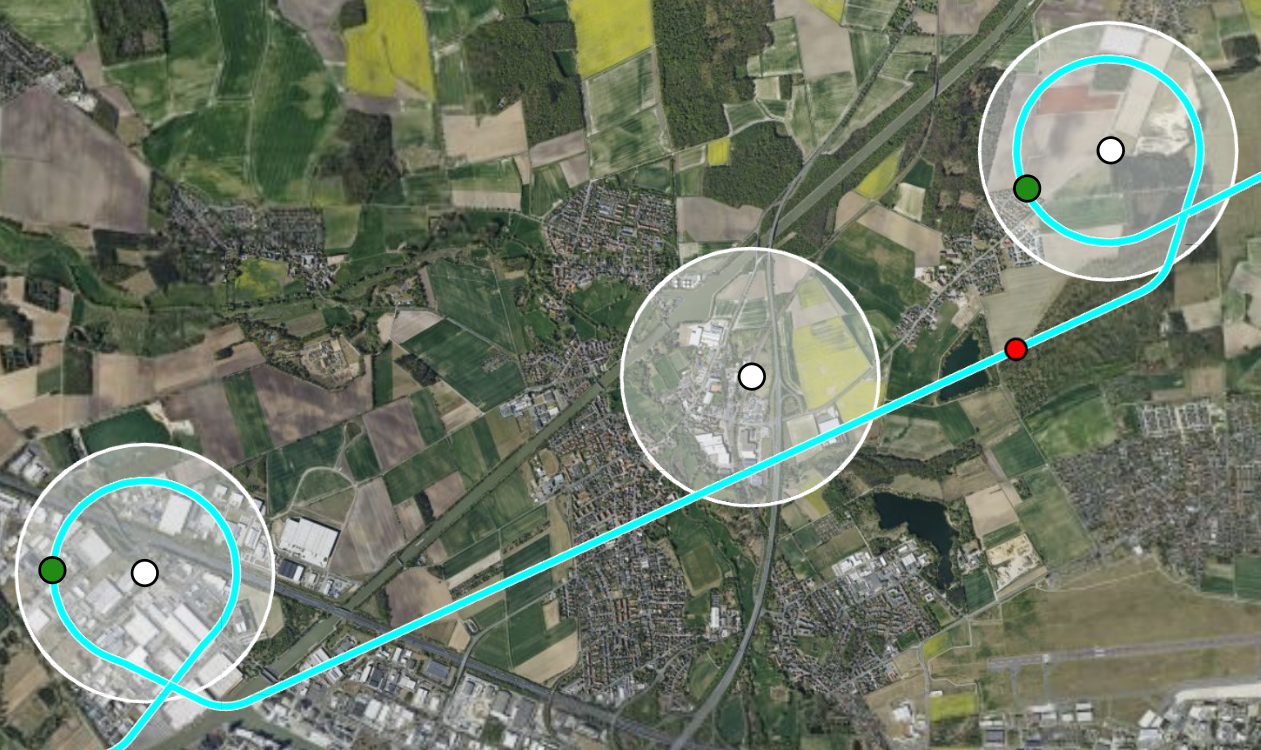} 
    \caption{An invalid flight}
    \end{subfigure}
    \caption{Traces of drones monitoring closeness to celltowers.}
    \label{fig:celltower_traces}
\end{figure}

\paragraph{Celltower Monitoring.}
As the first part of our evaluation, we consider a scenario in which a drone must periodically approach cell towers in order to transmit collected data.
To ensure transmission, the drone must remain sufficiently close to a tower continuously for two minutes.
The specification is satisfied if such a transmission period occurs at least once every four minutes.
Consider the following specification:
\begin{lstlisting}
input pos : (Float64, Float64)
input ct_res : (QueryId, UInt64) as (_, in_range_count)
output queries @1s@ :=
    query("NEARBY celltowers LIMIT 1 COUNT POINT ?1 ?2 700",
        with: pos.hold(or: (0.0, 0.0)), into: ct_res)
output in_range := in_range_count > 0
output in_range_s @1s@ := in_range.aggregate(over: 2min, using: forall)
trigger @1s@ $\neg$in_range_s.aggregate(over: 4min, using: exists)
\end{lstlisting}
The monitor periodically issues queries through the output stream \lstinline!queries! (Line~3), requesting the number of cell towers within a radius of 700 meter around the drone's current position.
Query results are received asynchronously via the input stream \lstinline!ct_res! (Line~2), from which the number of nearby towers is extracted as \lstinline!in_range_count!.
The stream \lstinline!in_range! (Line~6) evaluates whether at least one tower is currently within range.
Every second, \lstinline!in_range_s! (Line~7) checks that the drone has remained continuously within range throughout the preceding 2-minute interval using a sliding window aggregation.
Finally, the trigger (Line~8) verifies that at least one such uninterrupted interval has occurred within the last four minutes.

We generated synthetic flight traces that either satisfy or violate this property, as visualized in \Cref{fig:celltower_traces}.
Cell towers are marked by white dots, each surrounded by a circle representing the threshold for a drone to be considered nearby.
The points at which the drone was sufficiently long within this threshold are marked with a green dot, while violations are marked in red.
In the left trace, the drone circles around each tower, thereby satisfying the specification.
In contrast, the right trace omits circling one tower, resulting in a violation of the specification.

\paragraph{Powerline Monitoring.}
In this section, we compare the runtime of different database backends against an \rtlola baseline without external queries.
We implement an interface to SQLite3~\cite{sqlite3}, a general-purpose relational database system.
In addition, we integrate specialized backends for geospatial queries, including the Tile38 database~\cite{tile38}, as well as our own implementation based on k-d~trees.
A k-d~tree~\cite{DBLP:journals/cacm/Bentley75} is a data structure that organizes geospatial points in k-dimensional space, enabling fast nearest-neighbor queries by recursively subdividing the search space along alternating axes.
This evaluation shows that our interface generalizes across different backends, from general-purpose relational databases to specialized geospatial databases.

We evaluate all backends using a specification that monitors the distance to the nearest obstacle.
The full dataset consists of approximately 1.7 million powerline poles extracted from OpenStreetMap \footnote{\url{www.openstreetmap.org}}, with a small and medium subset of 10.000 and 100.000 poles, respectively.
As a baseline, we compare against a pure \rtlola implementation that uses parameterized streams to store and query all data internally.
The runtimes were obtained on a system with a 13th Gen Intel Core i7-1355U with 200 runs for the geospatial backends, and 20 for SQLite3 and the \rtlola baseline.

\begin{table}[t]
    \centering
    \caption{Overview of the evaluated query backends and their average runtime on a trace issuing 190 queries.}
    \label{tab:backends}
    \vspace*{2mm}
    \begin{tabular}{lllrrr}
        \toprule
        Backend & Purpose & Query Language & \multicolumn{3}{c}{Runtime} \\
        \cmidrule(lr){4-6}
         & & & Small & Medium & Full \\
        \midrule
        RTLola & Baseline & - & 876ms & 24.4s & - \\
        SQLite3~\cite{sqlite3} & General-purpose & SQL & 1.04s & 11.6s & 250s \\
        Tile38~\cite{tile38} & Geospatial & Tile38 Commands & 4.0ms & 3.9ms & 4.3ms \\
        Our kd-tree & Geospatial & Tile38 Commands & 2.8ms & 2.7ms & 3.0ms \\
        \bottomrule
    \end{tabular}
\end{table}

The runtimes for monitoring a trace with 190 queries are depicted in \Cref{tab:backends}.
For small datasets, the overhead of issuing external database queries dominates the runtime, leading to SQLite3 performing slightly worse than the pure \rtlola baseline.
However, the baseline scales poorly as the dataset size increases and is infeasible for the full dataset of 1.7 million entries, and is therefore omitted from the table.
SQLite3 scales slightly better, but still requires 250s for the full database, making it unsuitable for real-time monitoring.
In contrast, the geospatial backends achieve consistently low runtimes across all databases.
Our own implementation consistently outperforms the other backends and achieves the best overall runtime.

\begin{figure}[t]
    \centering
    \includegraphics[width=0.75\linewidth]{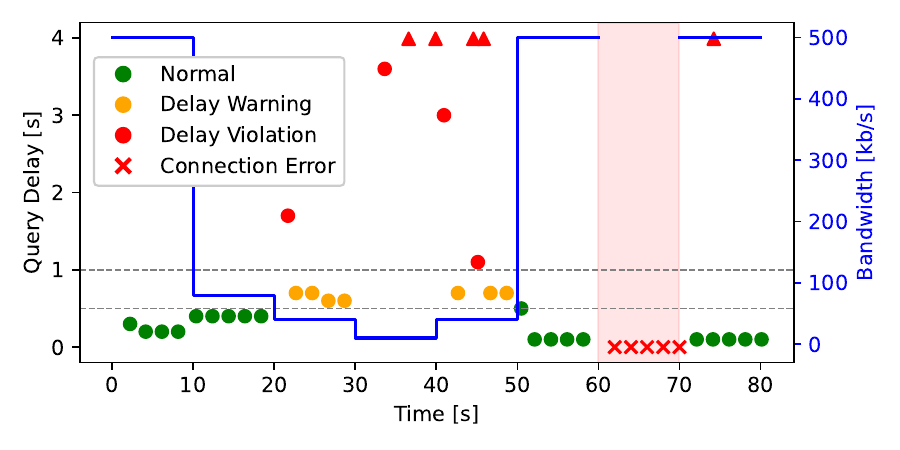}
    \caption{
        Plot showing the monitored query delay when querying a weather API with varying connection bandwidth.
    }
    \label{fig:weather_bandwidth}
\end{figure}

\paragraph{Weather-Aware Monitoring.}
As the final part of our evaluation, we monitor a dynamic backend in the form of a weather API.
In particular, we use the Bright Sky API~\cite{brightsky} to query the current temperature at the drone's position.
The specification tracks the delay between issuing a query and receiving the corresponding response.

The results of a drone flight are shown in \Cref{fig:weather_bandwidth}.
During the flight, we artificially vary the available bandwidth, which is depicted in blue in the graph, and at one point, the internet connection fails completely.
This interval is shaded red.
The scatter plot illustrates the issued triggers from the monitor throughout the flight.
Green circles indicate that a response arrived within the expected time bound (less than 0.5s), while orange markers denote warnings issued by the monitor for delays between 0.5s and 1.0s.
Red circles indicate delay violations exceeding 1.0s, and points exceeding the upper limit of the figure are represented by red triangles.
Finally, red crosses denote failed queries caused by the temporary loss of internet connectivity.
These results demonstrate that the monitor correctly detects degraded API performance, and gracefully handles complete connectivity loss without crashing.

\section{Conclusion}

We presented an extension to stream-based runtime monitoring that enables monitors to actively query external data sources at runtime.
Our approach ensures asynchronous and type-safe queries, allows specifications to handle query failures gracefully, and provides a backend-agnostic interface that abstracts away the complexities of integration into the monitor.
An evaluation on monitoring scenarios from the aviation domain confirms that the approach is practical and effective for applications that depend on both static and dynamic external data.

While our evaluation focused on the aviation domain, the approach is applicable to a wide range of systems that rely on external data sources.
An interesting direction for future work would be the integration of large language models as a type of backend, enabling monitors to reason about unstructured data such as natural language.

\begin{credits}
\subsubsection{\ackname} This work was partially supported by the German Research Foundation (DFG) as part of TRR 248 (No.~389792660) and PreCePT (No.~521273327), and by the European Research Council (ERC) Grant HYPER (No.~101055412).

\subsubsection{\discintname}
The authors have no competing interests to declare that are relevant to the content of this article.
\end{credits}

\bibliographystyle{splncs04}
\bibliography{bibliography.bib}

\ifthenelse{\boolean{fullversion}}{
    \clearpage
    \appendix
    \section{Specifications}\label{app:specs}

\subsection{Powerline Monitoring}

\noindent\textbf{RTLola Baseline}

\begin{lstlisting}
import math

input lat : Float64
input lon : Float64

input time_str : String

input powerline_lat : Float64
input powerline_lon : Float64

output powerline_distance(pl_lat, pl_lon)
  spawn with (powerline_lat, powerline_lon)
  eval @lat&&lon@ with sqrt((pl_lat - lat)**2.0 +  (pl_lon - lon)**2.0)

output closest_powerline := powerline_distance.aggregate(over_instances: fresh, using: min).defaults(to: 0.0)
\end{lstlisting}

\noindent\textbf{SQLite}

\begin{lstlisting}
import query

input close_powerlines : (QueryId, (Float64,Float64,Float64))
  as (query_id, (pl_lat,pl_lon,pl_distance))

#[query_status]
input query_status : (QueryId, QueryStatus) as (result_id, result_status)

input lat : Float64
input lon : Float64

input time_str : String

output query_close_powerlines := (time_str, query_unchecked("
SELECT
  lat,
  lon,
  ROUND(haversine(lat, lon, ?1, ?2), 1) as distance
  FROM static_objects
  ORDER BY distance
  LIMIT 1;
", with: (lat,lon), into: "close_powerlines"))
\end{lstlisting}

\noindent\textbf{Tile38}

\begin{lstlisting}
import query

input close_powerlines : (QueryId, (Float64,Float64,Float64))
  as (query_id, (pl_lat,pl_lon,pl_distance))

#[query_status]
input query_status : (QueryId, QueryStatus) as (result_id, result_status)

input lat : Float64
input lon : Float64

input time_str : String

output query_close_powerlines := (time_str, query_unchecked("NEARBY static_objects DISTANCE LIMIT 1 POINT ?1 ?2", with: (lat,lon), into: "close_powerlines"))
\end{lstlisting}

\subsection{Weather-Aware Monitoring}

\begin{lstlisting}
input lat : Float64
input lon : Float64
input time : Float64
#[query_status]
input status: (QueryId, QueryStatus) as (status_id, status_msg)
input temperature : (QueryId, Float64)
#[public]
output q @2s@ := query_unchecked(query: "", with: (lat.hold(or: 0.0), lon.hold(or: 0.0)), into: "temperature")
output q_time(qid)
  spawn with q
  eval when q == qid with time.hold(or: 0.0)
  close when status_id = qid
#[public]
output query_delay
  eval @status@ with time.hold(or: 0.0) - q_time(status_id).hold(or: 0.0)
trigger query_delay > 0.5 "slight delay"
trigger query_delay > 1.0 "long delay"
trigger status_msg == QueryStatus::Error "query error"
\end{lstlisting}
}{}

\end{document}